\documentclass[reprint,amsmath,amssymb,aps]{revtex4-2}
\usepackage{graphicx}
\usepackage{bm}
\usepackage{amsmath}
\usepackage{amssymb}
\usepackage{booktabs}
\usepackage{CJKutf8}
\usepackage{xcolor}
\usepackage{tikz}
\usepackage[colorlinks=true,linkcolor=blue,urlcolor=blue,citecolor=blue]{hyperref}

\definecolor{lime}{HTML}{A6CE39}
\DeclareRobustCommand{\orcidicon}{%
  \begin{tikzpicture}
  \draw[lime, fill=lime] (0,0) circle [radius=0.16]
    node[white] {{\fontfamily{qag}\selectfont \tiny ID}};
  \draw[white, fill=white] (-0.0625,0.095) circle [radius=0.007];
  \end{tikzpicture}%
  \hspace{-2mm}%
}
\foreach \x in {A, ..., Z}{%
  \expandafter\xdef\csname orcid\x\endcsname{%
    \noexpand\href{https://orcid.org/\csname orcidauthor\x\endcsname}{\noexpand\orcidicon}}
}

\begin{document}
\begin{CJK*}{UTF8}{gbsn}

\title{Constraining AGN Disk Properties with Gravitational Waves from Inspiraling Stellar-Mass Binary Black Holes in Hierarchical Triple Systems}

\author{Jie Wu (吴洁)\orcidA{}\hspace{0.7em}$^{1,2}$ }
\author{Mengfei Sun (孙孟飞)\orcidB{}\hspace{0.7em}$^{1,2}$ }
\author{Jin Li (李瑾)\orcidC{}\hspace{0.7em}$^{1,2,3}$ }
\email{cqujinli1983@cqu.edu.cn}
\author{Zhoujian Cao (曹周键)\orcidD{}\hspace{0.7em}$^{4,5,6}$ }
\email{zjcao@bnu.edu.cn}

\affiliation{$^{1}$College of Physics, Chongqing University, Chongqing 401331, China}
\affiliation{$^{2}$Department of Physics and Chongqing Key Laboratory for Strongly Coupled Physics, Chongqing University, Chongqing 401331, China}
\affiliation{$^{3}$Institute of Advanced Interdisciplinary Studies, Chongqing University, Chongqing 401331, China}
\affiliation{$^{4}$Department of Astronomy, Beijing Normal University, Beijing 100875, China}
\affiliation{$^{5}$Institute for Frontiers in Astronomy and Astrophysics, Beijing Normal University, Beijing 102206, China}
\affiliation{$^{6}$School of Fundamental Physics and Mathematical Sciences, Hangzhou Institute for Advanced Study, UCAS, Hangzhou 310024, China}

\begin{abstract}
Space-based gravitational-wave detectors can observe stellar-mass binary black holes (BBHs) long before merger, allowing weak environmental perturbations to accumulate.
For binaries embedded in active galactic nucleus (AGN) disks, the local gas density characterizes the environment of the supermassive black hole (SMBH) and compact-object migration.
We study whether such signals can constrain this density when a stellar-mass BBH orbits a Kerr SMBH.
We evolve the outer orbit with relativistic corrections and gaseous dynamical friction (DF), and construct the detector-frame waveform including BBH inspiral, de Sitter precession, DF phase correction, and moving-source effects.
Using Fisher-matrix calculations for sampled systems, we estimate statistical uncertainties and systematic errors.
Larger gas densities generally improve the statistical precision of several source and outer-orbit parameters, but also increase systematic errors when DF is omitted.
For favorable GW190521-like systems observed by LISA for one year, the disk density can be constrained at the level of $\sigma_{\rho}\sim10^{-12}\text{--}10^{-10}\,{\rm g\,cm^{-3}}$.
Such constraints would connect BBH merger environments to the gas structure of galactic nuclei and the conditions that support black hole growth.
These results indicate that hierarchical BBH inspirals can probe AGN disk environments, provided that gas effects are modeled consistently.
\end{abstract}
\maketitle
\end{CJK*}

\section{Introduction}\label{sec:introduction}
The detection of gravitational waves (GWs) from binary black holes (BBHs) has opened a direct observational window on compact-object dynamics in the strong-field regime~\cite{BBH_first_detection,GWTC4_catalog,GWTC5_catalog}.
Ground-based detectors observe the final seconds to minutes of BBH coalescences.
By contrast, space-based detectors such as LISA can observe stellar-mass BBHs in the millihertz band from a few months to thousands of years before merger~\cite{LISA_mission,LISA_stellar_origin_BBH}.
Such long baselines allow weak, slowly varying perturbations to accumulate into measurable phase and amplitude modulations~\cite{Environmental_effects_GW,SMBH_triple_acceleration,SMBH_triple_mass_redshift}.
Inspiraling stellar-mass BBHs can therefore probe both compact-object properties and their astrophysical environments.

Active galactic nucleus (AGN) disks provide a particularly relevant environment for such probes.
The dense gaseous disk around a supermassive black hole (SMBH) can facilitate black hole (BH) capture, migration, binary hardening, and repeated mergers~\cite{AGN_BBH_factories,AGN_disk_migration_traps,AGN_eccentric_BBH_factories}.
This connection links compact-object dynamics to the structure of the central accretion flow.
The event GW190521 has further motivated this channel.
Its large total mass and proposed electromagnetic counterpart S190521g have been discussed as possible evidence for an AGN-disk merger, although this interpretation remains model dependent~\cite{GW190521_detection,GW190521_properties,GW190521_AGN_counterpart}.
Even without an electromagnetic counterpart, a BBH embedded in or near an AGN disk may carry environmental information in its GW signal.

Several waveform ingredients for hierarchical systems near SMBHs have already been developed.
Inayoshi \textit{et al.} showed that the center-of-mass acceleration of a BBH can leave an observable imprint on the GW phase~\cite{SMBH_triple_acceleration}.
Sources near SMBHs are also affected by the mass-redshift degeneracy~\cite{SMBH_triple_mass_redshift}.
Yu and Chen demonstrated that de Sitter-like (dS) precession and the Doppler shift can help determine SMBH properties, providing useful guidance for extracting information from hierarchical triple systems~\cite{SMBH_triple_direct_measurement}.
Spin-dependent orbital effects and the transverse Doppler effect provide additional corrections~\cite{SMBH_triple_spin,SMBH_triple_transverse_Doppler}.
Waveform calculations for moving sources further show that the rest-frame signal must be transformed to the detector frame when the source velocity changes in magnitude or direction~\cite{Moving_source_waveform,Accelerating_source_waveform}.
These studies indicate that hierarchical BBH systems can encode dynamical information absent from isolated binaries.

Gas effects provide another route by which the environment can enter GW signals.
For GW190521-like binaries that could be observed by future space-based detectors such as LISA, environmental corrections can become detectable, and dynamical friction (DF) can bias parameter estimation if neglected~\cite{GW190521_LISA_environment,GW190521_LISA_orbit_environment,Accretion_physics_GW}.
Recent work has also explored direct inference of AGN disk density profiles from GW data and found that disk parameters can be constrained in favorable source configurations~\cite{AGN_disk_profile_GW}.
Gas DF is the gravitational drag produced by the density wake of a massive perturber moving through gas~\cite{Gas_dynamical_friction}.
For an object moving in a gaseous disk, the drag can have both azimuthal and radial components.
Its direction depends on the relative velocity between the perturber and the local disk flow~\cite{Circular_orbit_GDF,Eccentric_orbit_GDF}.
These considerations motivate a model that treats the gas DF force, relativistic outer orbit, inner-binary precession, and moving-source transformation consistently.
Such a framework is required to connect waveform imprints to AGN disk properties.

Building on our previous work on moving-source waveform effects and hierarchical BBH probes~\cite{my_paper_smbh,my_paper_gc,my_paper_kick}, we study LISA observations of stellar-mass BBHs in hierarchical AGN disk systems.
We evolve the outer orbit with high-order post-Newtonian (PN) terms and gaseous DF.
The rest-frame signal is described by a time-domain 3.5PN waveform including dS precession and the DF phase correction.
We then apply the moving-source transformation to obtain the waveform of a BBH orbiting the SMBH.
Finally, we use the Fisher information matrix (FIM) to estimate parameter uncertainties and systematic errors.
This framework assesses when AGN gas effects are measurable and when they must be included in precision inference.

The paper is organized as follows.
Section~\ref{sec:triple_system} describes the hierarchical triple configuration, AGN disk model, and DF prescription.
Section~\ref{sec:gw_signal} presents the GW signal model and the moving-source transformation.
Section~\ref{sec:methodology} gives the parameter-estimation method and simulation setup.
Section~\ref{sec:results} discusses the impact of DF on source parameters and the constraints on AGN disk properties.
Section~\ref{sec:conclusions} summarizes the main conclusions.
Throughout this paper, we use geometrized units $G=c=1$ unless otherwise stated.

\section{Hierarchical triple in an AGN disk}\label{sec:triple_system}

\begin{figure*}[t]
    \centering
    \includegraphics[width=0.95\textwidth]{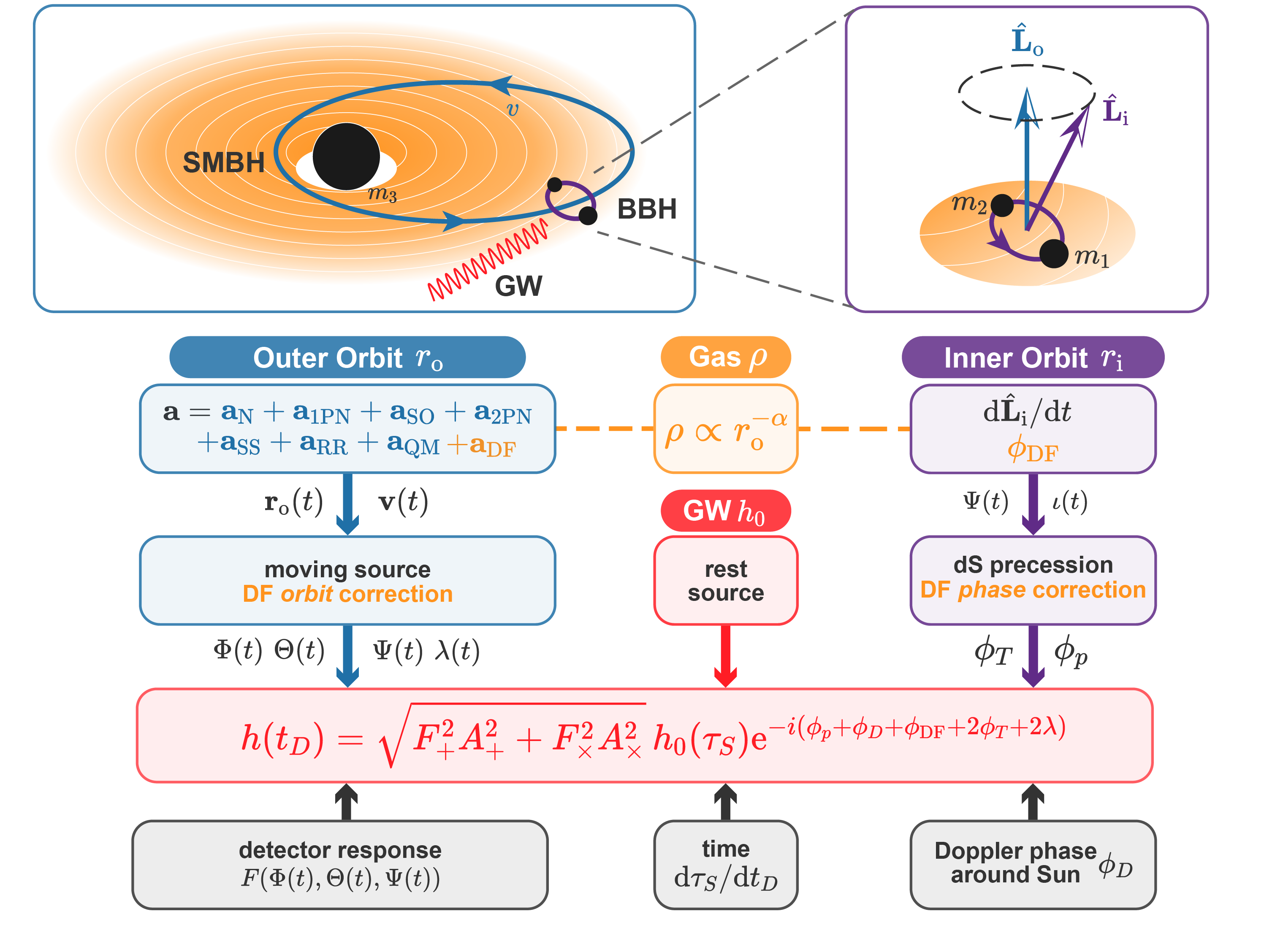}
    \caption{
    Schematic of the hierarchical triple system in an AGN disk and of the waveform-modeling flow.
    The upper left panel shows the stellar-mass BBH orbiting the SMBH as the outer binary, while the upper right panel enlarges the inner BBH orbit.
    The lower flow chart summarizes the signal construction.
    The outer orbit supplies $\mathbf r_{\rm o}(t)$ and $\mathbf v(t)$ for the moving-source transformation, the AGN disk density determines the DF effect, and the inner binary supplies the rest-frame inspiral waveform with dS precession and the DF phase correction.
    }
    \label{fig:AGN}
\end{figure*}

We consider a hierarchical triple in which a stellar-mass BBH orbits a central SMBH inside an AGN disk.
The two stellar-mass BHs, with masses $m_1$ and $m_2$, form the inner binary.
The BBH center of mass and the SMBH of mass $m_3$ form the outer binary.
Subscripts ``i'' and ``o'' denote inner- and outer-orbit quantities when needed.
The upper part of Fig.~\ref{fig:AGN} illustrates this configuration.
It shows the outer SMBH--BBH orbit on the left and an enlarged view of the inner BBH orbit on the right.
For $m_3\gg m_1+m_2$ and semimajor axes satisfying $a_{\rm o}\gg a_{\rm i}$, the inner and outer motions remain dynamically separated.
The system can therefore be treated as a stable hierarchical triple~\cite{Hierarchical_triple_stability}. Such systems can be subject to several secular effects. For example, Lidov--Kozai (LK) oscillations periodically modulate the inner-orbit eccentricity and the mutual inclination of the inner and outer orbital planes~\cite{Lidov_Kozai_binaries_MBH}. The characteristic LK period is~\cite{SMBH_triple_direct_measurement}
\begin{equation}
\begin{aligned}
P_{\rm LK}&\simeq1.2\times10^3{\rm yr}(1-e_{\rm o}^2)^{3/2}\\
&\quad\times\left(\frac{m_3}{10^8 \rm M_\odot}\right)^2
\left(\frac{a_{\rm o}}{100R_s}\right)^3
\left(\frac{f_{\rm GW}}{1{\rm mHz}}\right).
\end{aligned}
\label{eq:LK_period}
\end{equation}
For the fiducial parameters, the LK timescale is of order $10^3\,{\rm yr}$. This quantity is a characteristic quadrupole LK timescale rather than an exact oscillation period for arbitrary initial conditions, and it can decrease for an eccentric outer orbit~\cite{Antognini2015}. The corresponding closest distance of the BBH center of mass from the SMBH can therefore enter a stronger-field region. For the final effective sample, the inner-binary GR-precession timescale is shorter than the LK timescale for the great majority of sources, so rapid GR precession suppresses coherent LK eccentricity growth~\cite{Naoz2016} and supports the reliability of the reported trends within the adopted baseline model. We nevertheless do not include secular LK evolution in the waveform, and regard sources with short LK timescales or close approaches to the SMBH as outside the regime in which this approximation can be assumed without qualification. Further validity conditions for the hierarchical treatment are discussed in Ref.~\cite{my_paper_smbh}.

For the outer-orbit calculation, we model the central SMBH as a spinning Kerr BH.
The BBH center of mass follows a relativistic orbit around the SMBH.
This orbit provides the time-dependent source position and velocity required by the moving-source waveform transformation~\cite{SMBH_triple_spin,Moving_source_waveform}.
Because gas DF continuously perturbs the outer motion, we solve the orbit numerically.
We do not approximate the BBH center-of-mass motion by a constant acceleration or a fixed Keplerian orbit.
The lower part of Fig.~\ref{fig:AGN} summarizes how the outer orbit, gas density, inner-binary waveform, detector response, and moving-source transformation combine to produce the final GW signal.

The outer-orbit position satisfies the second-order equation $\ddot{\mathbf r}_{\rm o}=\mathbf a$.
For numerical integration, we rewrite this equation as the first-order system
\begin{equation}
    \frac{{\rm d}}{{\rm d}t}
    \begin{pmatrix}
        \mathbf r_{\rm o} \\
        \mathbf v
    \end{pmatrix}
    =
    \begin{pmatrix}
        \mathbf v \\
        \mathbf a(\mathbf r_{\rm o},\mathbf v)
    \end{pmatrix}.
    \label{eq:outer_first_order_system}
\end{equation}
The total acceleration is
\begin{equation}
    \mathbf a=\mathbf a_{\rm DF}+\mathbf a_{\rm PN}.
    \label{eq:outer_acceleration}
\end{equation}
Here $\mathbf a_{\rm DF}$ is the gaseous DF contribution, while $\mathbf a_{\rm PN}$ describes the vacuum orbital dynamics around the spinning SMBH in a PN expansion.
The PN acceleration can be expanded as~\cite{PN_spin_terms,PN_quadrupole_monopole}
\begin{equation}
    \mathbf a_{\rm PN}
    =
    \mathbf a_{\rm N}
    +\mathbf a_{\rm 1PN}
    +\mathbf a_{\rm SO}
    +\mathbf a_{\rm 2PN}
    +\mathbf a_{\rm SS}
    +\mathbf a_{\rm RR}
    +\mathbf a_{\rm QM}.
    \label{eq:pn_decomposition}
\end{equation}
The term $\mathbf a_{\rm N}$ is the Newtonian acceleration.
The terms $\mathbf a_{\rm 1PN}$ and $\mathbf a_{\rm 2PN}$ are the conservative 1PN and 2PN corrections.
$\mathbf a_{\rm SO}$ and $\mathbf a_{\rm SS}$ are the spin-orbit and spin-spin terms, $\mathbf a_{\rm RR}$ is the radiation-reaction term, and $\mathbf a_{\rm QM}$ is the quadrupole-monopole interaction term.
The explicit expressions for these PN terms are given in Refs.~\cite{PN_spin_terms,PN_quadrupole_monopole}.
We generate the initial conditions from a Keplerian orbit.
We then numerically integrate Eqs.~\eqref{eq:outer_first_order_system} and \eqref{eq:outer_acceleration} to obtain $\mathbf r_{\rm o}(t)$ and $\mathbf v(t)$.
Compared with a purely Keplerian outer orbit, this treatment includes higher-order relativistic corrections and the environmental perturbation from AGN gas DF.

\begin{figure*}[t]
    \centering
    \includegraphics[width=0.95\textwidth]{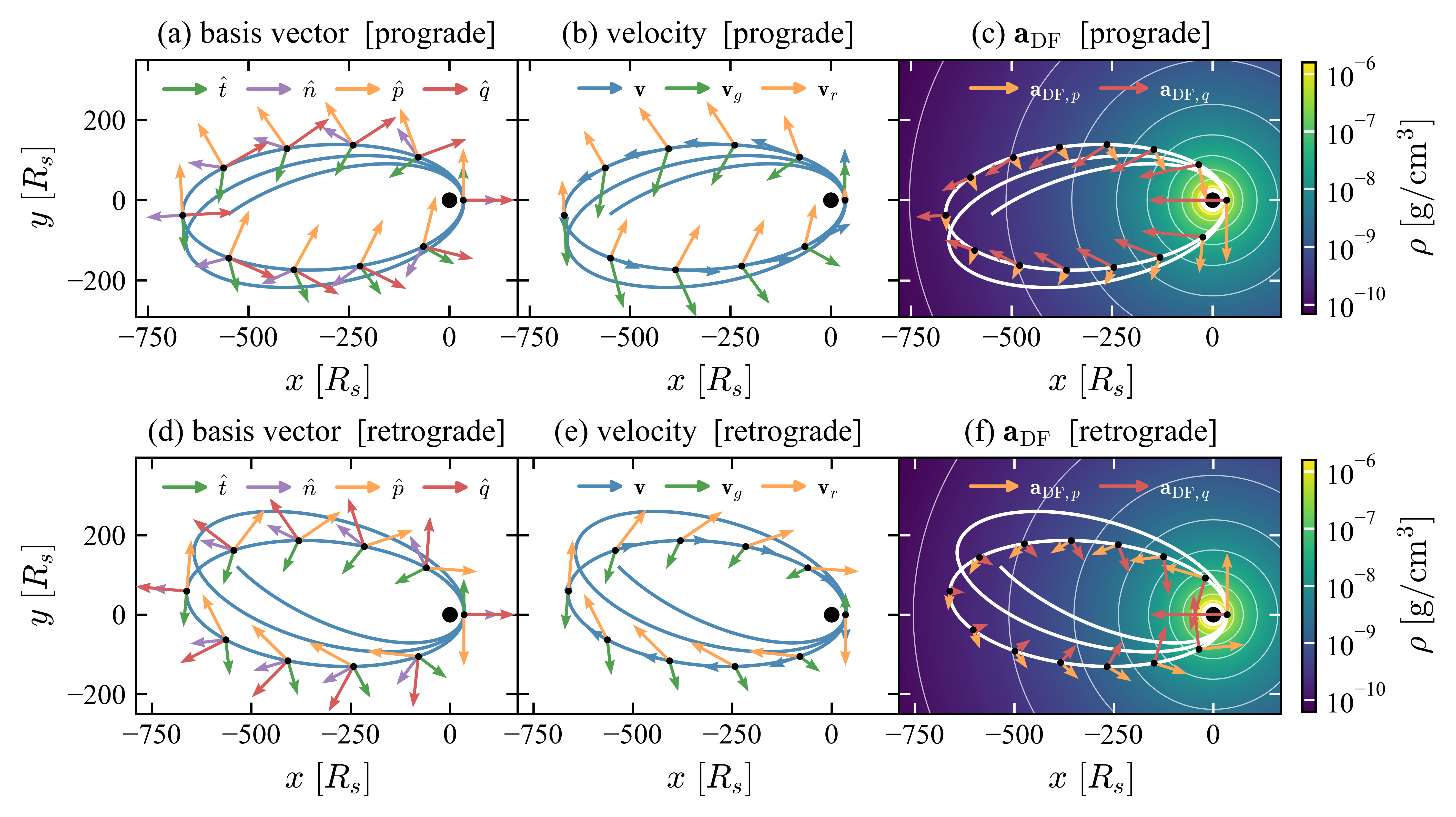}
    \caption{
    Comparison of the relevant vector components along the outer orbit over a 15-day interval.
    The upper and lower rows correspond to prograde and retrograde configurations, respectively.
    The parameters are $m_3=10^6\,M_\odot$, $a_{\rm o}=350\,R_s$, $e_{\rm o}=0.9$, $\chi_3=0.9$, and $\rho_0=10^{-9}\,{\rm g\,cm^{-3}}$.
    Panels (a) and (d) show the directions of the $(\hat{\mathbf t},\hat{\mathbf n})$ and $(\hat{\mathbf p},\hat{\mathbf q})$ bases along the orbit.
    The different arrow lengths are used only for visual separation and do not represent different unit-vector magnitudes.
    Panels (b) and (e) show the BBH center-of-mass velocity, gas velocity, and relative velocity.
    At each point, the arrow lengths indicate the relative magnitudes of the three velocities.
    Arrows at different orbital points are not plotted on a common scale.
    Panels (c) and (f) show the $\mathbf a_{\rm DF,p}$ and $\mathbf a_{\rm DF,q}$ components in the $(\hat{\mathbf p},\hat{\mathbf q})$ basis.
    The two component arrows are not on the same scale, and point-to-point arrow lengths are plotted logarithmically.
    The background color gives the gas density on a logarithmic scale.
    }
    \label{fig:orbit_component}
\end{figure*}
For the AGN gas, we adopt a power-law density profile motivated by parametrized disk descriptions used in GW environmental studies~\cite{Accretion_physics_GW,AGN_disk_profile_GW},
\begin{equation}
    \rho(r_{\rm o})=\rho_0\left(\frac{r_{\rm o}}{r_{\rm ref}}\right)^{-\alpha},
    \label{eq:disk_density}
\end{equation}
where $R_s=2m_3$ is the Schwarzschild radius of the SMBH, $\alpha$ is the radial density index, and $r_{\rm ref}$ is a reference radius.
We set $r_{\rm ref}=350\,R_s$, with density normalization $\rho_0=\rho(350\,R_s)$.
The two parameters $\rho_0$ and $\alpha$ therefore describe the gas distribution used in this work.
This power-law profile is a simplified local description of the AGN disk over the radial region sampled by the BBH center of mass.
Although this profile does not replace a full accretion-disk solution, it provides a compact parametrization for the outer-orbit calculation and FIM analysis.

The density profile sets the strength of the gas response, while the direction of the DF force is determined by the BBH center-of-mass motion relative to the local disk flow.
We assume that the gas follows a Keplerian azimuthal flow~\cite{Accretion_physics_GW,AGN_disk_profile_GW},
\begin{equation}
    v_K=\sqrt{\frac{m_3}{r_{\rm o}}},\qquad
    \mathbf v_{\rm g}=v_K\hat{\mathbf t},
    \label{eq:gas_velocity}
\end{equation}
where
\begin{equation}
    \hat{\mathbf n}=\frac{\mathbf r_{\rm o}}{r_{\rm o}},
    \qquad
    \hat{\mathbf t}=\hat{\mathbf L}_{\rm o}\times\hat{\mathbf n}
    \label{eq:radial_tangential_basis}
\end{equation}
are the radial and tangential unit vectors in the outer-orbit plane.
Here $\hat{\mathbf L}_{\rm o}$ is the unit angular-momentum vector of the outer orbit.
The sound speed is set by $c_s=h v_K$, with disk aspect ratio $h\simeq0.01$.

Our DF prescription is based primarily on the work of Kim and Kim (KK07), which calculated gaseous DF on a perturber in a circular orbit through a uniform medium~\cite{Circular_orbit_GDF}.
We also use the later extension to elliptical Keplerian orbits~\cite{Eccentric_orbit_GDF}.
KK07 showed that a curved gaseous wake produces both an azimuthal component opposite to the direction of motion and a radial component.
The radial component contributes little to orbital decay for circular motion, but it can become relevant for noncircular orbital dynamics.
In an AGN disk, the gas rotates rather than remaining at rest, and the BBH center-of-mass orbit is not exactly circular.
The DF force should therefore be determined by the velocity relative to the local gas, rather than by the absolute BBH velocity.
We define
\begin{equation}
    \mathbf v_{\rm r}=\mathbf v-\mathbf v_{\rm g}.
    \label{eq:relative_velocity}
\end{equation}
This relative velocity defines the local basis in which we express the DF force.
The unit vector along the relative gas velocity is
\begin{equation}
    \hat{\mathbf p}=\frac{\mathbf v_{\rm r}}{v_{\rm r}},
    \label{eq:p_unit_vector}
\end{equation}
and its projection onto the outer-orbit radial-tangential basis is
\begin{equation}
    p_R=\hat{\mathbf p}\cdot\hat{\mathbf n},
    \qquad
    p_\phi=\hat{\mathbf p}\cdot\hat{\mathbf t}.
    \label{eq:p_components}
\end{equation}
Thus
\begin{equation}
    \hat{\mathbf p}=p_R\hat{\mathbf n}+p_\phi\hat{\mathbf t}.
    \label{eq:p_decomposition}
\end{equation}
For the DF component perpendicular to the relative velocity, we define
\begin{equation}
    \hat{\mathbf q}=s_{\rm o}(p_\phi\hat{\mathbf n}-p_R\hat{\mathbf t}),
    \label{eq:q_basis}
\end{equation}
where $s_{\rm o}=+1$ and $s_{\rm o}=-1$ correspond to prograde and retrograde outer orbits, respectively.
This sign choice keeps the orientation of $\hat{\mathbf q}$ consistent for opposite orbital senses and satisfies $\hat{\mathbf p}\cdot\hat{\mathbf q}=0$.
Panels (a) and (d) of Fig.~\ref{fig:orbit_component} show the two bases along the orbit.
Panels (b) and (e) show that $\hat{\mathbf p}$ follows $\mathbf v_{\rm r}$, while $\hat{\mathbf t}$ follows the local gas velocity, as implied by Eqs.~\eqref{eq:gas_velocity} and \eqref{eq:p_unit_vector}.
We use the KK07 prescription as a local, quasi-steady DF extension rather than as a global solution for the entire AGN disk. At each outer-orbit integration step, we evaluate the density, sound speed, and gas velocity at the instantaneous radius. We then transform to the local gas rest frame and construct the instantaneous $(\hat{\mathbf p},\hat{\mathbf q})$ basis from the relative velocity defined above. The KK07 azimuthal and radial wake components are evaluated from the instantaneous Mach number and projected onto this basis. This procedure recovers the KK07 force in the circular, uniform-medium limit. For eccentric trajectories in a radially structured disk, it is a local circular-proxy approximation. Wake-history effects and global disk structure are not included.

This prescription retains the leading dependence of the DF force on local disk conditions while omitting vertical disk structure, turbulence, perturber feedback, and nonlinear or nonlocal wake evolution. DF modeling for hierarchical triples embedded in AGN disks remains under active development. The local model nevertheless provides a tractable way to incorporate the leading gas effect into the waveform calculation.

\begin{figure}[!ht]
    \centering
    \includegraphics[width=0.97\columnwidth]{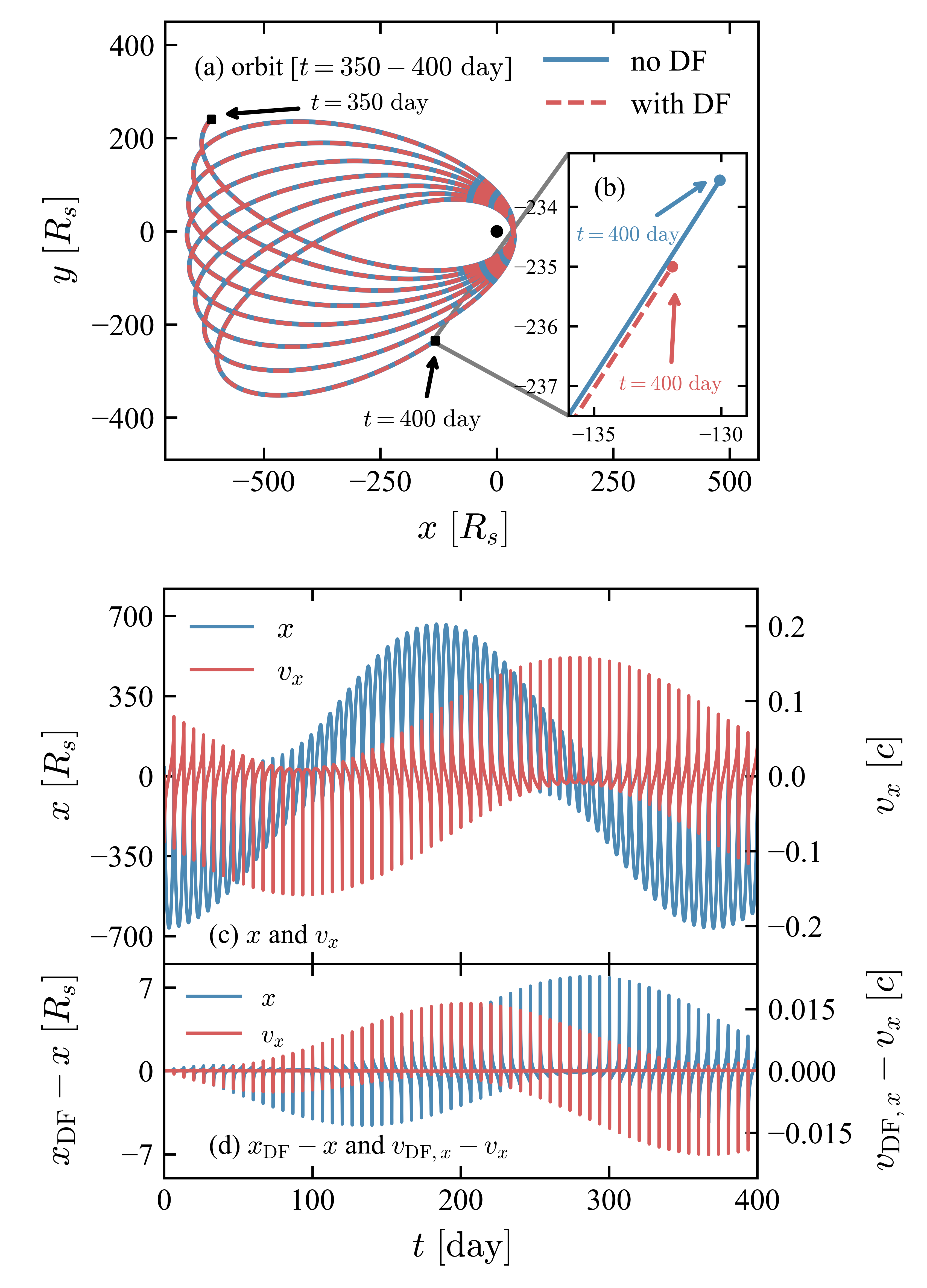}
    \caption{
    Outer orbit with and without gas DF for the same parameters as Fig.~\ref{fig:orbit_component}.
    Panel (a) shows the orbit from $t=350$ to $400\,{\rm day}$.
    Panel (b) zooms in on the last half hour before $t=400\,{\rm day}$.
    Panel (c) shows the time evolution of the $x$ component of the position and velocity.
    Panel (d) shows the corresponding differences $x_{\rm DF}-x$ and $v_{{\rm DF},x}-v_x$ between the DF and no-DF orbits.
    }
    \label{fig:orbit_DF}
\end{figure}

In the $(\hat{\mathbf p},\hat{\mathbf q})$ basis, the gas DF acceleration is written as
\begin{equation}
    \mathbf a_{\rm DF}
    =\mathcal A_{\rm DF}
    \left[
        I_\phi(\mathcal M)\hat{\mathbf p}
        +
        I_R(\mathcal M)\hat{\mathbf q}
    \right],
    \label{eq:df_acceleration}
\end{equation}
where
\begin{equation}
    \mathcal A_{\rm DF}=-\frac{4\pi m\rho}{v_{\rm r}^2}.
    \label{eq:df_amplitude}
\end{equation}
Here $\mathcal M=v_{\rm r}/c_s$ is the Mach number, and $m=m_1+m_2$ is the total BBH mass.
The coefficient $I_\phi(\mathcal M)$ gives the dissipative component opposite to the relative motion and mainly controls angular-momentum loss.
The coefficient $I_R(\mathcal M)$ gives the transverse correction produced by the curved gaseous wake and affects the radial motion and eccentricity evolution~\cite{Circular_orbit_GDF,Eccentric_orbit_GDF}.
The transformation between the $(\hat{\mathbf t},\hat{\mathbf n})$ and $(\hat{\mathbf p},\hat{\mathbf q})$ bases does not introduce new physics.
It only maps the DF decomposition, which is defined relative to the perturber's motion through gas, into the rotating-disk and noncircular-orbit geometry used in our orbital integration.
In the systems studied here, the Mach number is typically $\mathcal M\gg4.4$.
According to the KK07 fitting formula, $I_R$ can be much larger than $I_\phi$ in this high-Mach regime.
However, the component that most directly removes orbital angular momentum is $I_\phi$~\cite{Circular_orbit_GDF}.
Panels (c) and (f) of Fig.~\ref{fig:orbit_component} should therefore be interpreted as a directional decomposition, not as a linear comparison of the two component magnitudes.
In those panels, the $\mathbf a_{\rm DF,p}$ component points opposite to $\hat{\mathbf p}$, while the $\mathbf a_{\rm DF,q}$ component is transverse to the relative motion.

Figure~\ref{fig:orbit_DF} illustrates how DF accumulates continuously in the outer orbit.
Panels (a) and (b) show that the DF-evolved orbit separates from the no-DF orbit over time, even though the instantaneous DF acceleration is weak.
The time-domain differences in panels (c) and (d) contain rapid oscillations on the outer orbital period and a slower envelope associated with periapsis precession.
The velocity difference is especially important for the GW signal because the moving-source transformation depends directly on $\mathbf v(t)$.
The system in Figs.~\ref{fig:orbit_component} and~\ref{fig:orbit_DF} is chosen to make the accumulated DF effect visually clear.
In part of the parameter space considered below, the DF-induced orbital difference is intrinsically much smaller.
Different AGN disk parameters produce different DF effects.
The accumulated velocity modulation therefore provides an indirect way to constrain the gas density profile.
The next section shows how this outer-orbit information enters the observed waveform.

\section{Gravitational-wave signal}\label{sec:gw_signal}

The waveform is constructed according to the flow summarized in the lower part of Fig.~\ref{fig:AGN}.
The inner BBH supplies the rest-frame inspiral signal, while the gas density sets the DF phase correction.
The dS effect changes the orientation of the inner binary.
The outer orbit supplies $\mathbf r_{\rm o}$ and $\mathbf v$ for the moving-source transformation.
This combination connects the AGN disk model and the hierarchical outer orbit to the signal measured by a space-based detector.

In the BBH rest frame, we model the inspiral signal with a 3.5PN+SO+SS time-domain waveform for a quasicircular inner binary~\cite{my_paper_PN_models}.
We use this model because it provides a time-domain waveform with spin effects while keeping the repeated waveform evaluations required by the FIM computationally tractable.
This PN model describes the long inspiral stage in the millihertz band, where the slowly evolving BBH can be observed for years before merger.
The rest-frame GW signal is then written as
\begin{equation}
    h_0=\mathcal A {\rm e}^{i\phi_{\rm PN}},
\end{equation}
with the amplitude
\begin{equation}
    \mathcal A=\frac{4}{D_L}M_c^{5/3}(\pi f)^{2/3}.
\end{equation}
Here $\phi_{\rm PN}$ is the PN phase, $M_c=\eta^{3/5}m$, and $\eta=m_1m_2/m^2$.
The explicit expansion of $\phi_{\rm PN}$ follows the standard nonspin, spin-orbit, and spin-spin contributions given in Refs.~\cite{PN_inspiral_review,PN_SO_review,PN_SS_effects}.
Environmental and three-body effects then enter as additional phase, orientation, response, and time-transformation effects on this rest waveform.

Gas DF also modifies the intrinsic phase of the inner BBH. Constant-density frequency-domain DF corrections have been used for quasicircular binaries~\cite{GW190521_LISA_environment,AGN_disk_profile_GW}. Here, we instead evaluate the inner-binary contribution on the time grid of the eccentric outer orbit. At each source-time sample, the outer-orbit solution provides the local disk state $\rho(t)$ and $c_s(t)$. We define the total and reduced masses as $m=m_1+m_2$ and $\mu=m_1m_2/m$. For a quasicircular inner orbit, the relative velocity has magnitude $v_{\rm i}=(\pi m f)^{1/3}$ and can be written as $\mathbf v_{\rm i}=v_{\rm i}\hat{\mathbf t}_{\rm i}(\varphi)$, where $\varphi$ is the inner-orbit phase and $\hat{\mathbf t}_{\rm i}$ is the instantaneous tangential unit vector. The component velocities in the BBH center-of-mass frame are
\begin{equation}
    \mathbf u_1=\frac{m_2}{m}\mathbf v_{\rm i},
    \qquad
    \mathbf u_2=-\frac{m_1}{m}\mathbf v_{\rm i}.
    \label{eq:inner_component_velocities}
\end{equation}
We evaluate the DF acceleration of each component using the local KK07 prescription in Eq.~(\ref{eq:df_acceleration}). The instantaneous velocity relative to the gas is
\begin{equation}
    \mathbf w_i
    =\mathbf u_i+\mathbf v-\mathbf v_{\rm g}
    =\mathbf u_i+\mathbf v_{\rm r},
    \label{eq:inner_component_gas_velocity}
\end{equation}
where $i=1,2$. We obtain the corresponding internal-energy change by averaging the differential DF work over the fast inner orbit at fixed outer-orbit time,
\begin{equation}
    \dot E_{\rm DF}(t,f)=
    \mu\left\langle
    \mathbf v_{\rm i}\cdot
    \left(\mathbf a_{{\rm DF},1}-\mathbf a_{{\rm DF},2}\right)
    \right\rangle_{\varphi},
    \label{eq:inner_df_energy_flux}
\end{equation}
thereby removing the common acceleration of the BBH center of mass. The Keplerian energy of the circular inner orbit is
\begin{equation}
    E_{\rm orb}
    =-\frac{1}{2}\mu(\pi m f)^{2/3}.
    \label{eq:inner_df_orbital_energy}
\end{equation}
Gaseous DF changes the inner-orbit energy at the rate $\dot E_{\rm DF}$. Since the orbital state is parametrized by the instantaneous frequency $f$, the energy change obeys
\begin{equation}
    \dot E_{\rm DF}
    =\frac{{\rm d}E_{\rm orb}}{{\rm d}f}\dot f_{\rm DF}.
    \label{eq:inner_df_energy_balance}
\end{equation}
Using Eqs.~(\ref{eq:inner_df_orbital_energy}) and (\ref{eq:inner_df_energy_balance}), we obtain the DF-induced frequency evolution
\begin{equation}
    \dot f_{\rm DF}
    =-\frac{3f^{1/3}}{\mu(\pi m)^{2/3}}\dot E_{\rm DF}.
    \label{eq:inner_df_frequency}
\end{equation}
The total chirp rate is the sum of the vacuum and DF-induced contributions,
\begin{equation}
    \dot f=\dot f_0(f)+\dot f_{\rm DF}(f),
    \label{eq:inner_df_total_frequency}
\end{equation}
where $\dot f_0(f)$ is the vacuum chirp rate. We write $f(t)=f_0(t)+\delta f(t)$, where $f_0(t)$ denotes the vacuum frequency trajectory and $\delta f(t)$ the DF-induced perturbation. Its time derivative is
\begin{equation}
    \dot f
    =\frac{{\rm d}f_0}{{\rm d}t}+\frac{{\rm d}\,\delta f}{{\rm d}t}
    =\dot f_0(f_0)+\frac{{\rm d}\,\delta f}{{\rm d}t}.
    \label{eq:inner_df_frequency_decomposition}
\end{equation}
Equation~(\ref{eq:inner_df_total_frequency}) then becomes
\begin{equation}
    \dot f
    =\dot f_0(f_0+\delta f)
    +\dot f_{\rm DF}(f_0+\delta f).
    \label{eq:inner_df_frequency_actual}
\end{equation}
Combining these two expressions gives
\begin{equation}
    \dot f_0(f_0)+\frac{{\rm d}\,\delta f}{{\rm d}t}
    =\dot f_0(f_0+\delta f)
    +\dot f_{\rm DF}(f_0+\delta f).
    \label{eq:inner_df_perturbation_substitution}
\end{equation}
The vacuum chirp rate on the right-hand side is expanded about the vacuum trajectory as
\begin{equation}
    \dot f_0(f_0+\delta f)
    \simeq
    \dot f_0(f_0)
    +\left.\frac{{\rm d}\dot f_0}{{\rm d}f}\right|_{f_0}\delta f.
    \label{eq:inner_df_chirp_expansion}
\end{equation}
Both $\delta f$ and $\dot f_{\rm DF}$ are first order in the environmental perturbation. The DF chirp rate therefore expands as
\begin{equation}
    \dot f_{\rm DF}(f_0+\delta f)
    =\dot f_{\rm DF}(f_0)
    +\left.\frac{{\rm d}\dot f_{\rm DF}}{{\rm d}f}\right|_{f_0}\delta f
    \simeq\dot f_{\rm DF}(f_0).
    \label{eq:inner_df_perturbation_order}
\end{equation}
The term proportional to $\delta f$ in Eq.~(\ref{eq:inner_df_perturbation_order}) is second order and is neglected. Combining Eqs.~(\ref{eq:inner_df_perturbation_substitution})--(\ref{eq:inner_df_perturbation_order}) gives
\begin{equation}
    \frac{{\rm d}\,\delta f}{{\rm d}t}
    =
    \left.\frac{{\rm d}\dot f_0}{{\rm d}f}\right|_{f_0}\delta f
    +\dot f_{\rm DF}(f_0).
    \label{eq:inner_df_phase_evolution}
\end{equation}
Solving Eq.~(\ref{eq:inner_df_phase_evolution}) for the DF-induced frequency perturbation $\delta f$ gives the corresponding GW phase correction $\phi_{\rm DF}$ through
\begin{equation}
    \frac{{\rm d}\phi_{\rm DF}}{{\rm d}t}=-2\pi\delta f.
    \label{eq:inner_df_phase_correction}
\end{equation}
with $\delta f(t_0)=\phi_{\rm DF}(t_0)=0$. For the vacuum waveform, the PN phase satisfies $\dot\phi_{\rm PN}=2\pi f_0$. The local density, sound speed, and BBH--gas relative velocity are therefore evaluated consistently along the eccentric outer orbit. In the high-Mach regime considered here, the DF force decreases as the gas-relative velocity increases. Since an eccentric outer orbit generally gives $\mathbf v_{\rm r}\ne0$, the component velocities relevant for DF cannot be replaced by those in the BBH center-of-mass frame. A circular comoving prescription with $\mathbf v\simeq\mathbf v_{\rm g}$ would omit this effect and can overestimate the DF contribution.

We next include the secular reorientation of the inner binary.
The dS effect drives precession of the inner orbital angular momentum~\cite{SMBH_triple_direct_measurement,SMBH_triple_eccentric_orbit},
\begin{equation}
    \frac{{\rm d}\hat{\mathbf L}_{\rm i}}{{\rm d}t}
    =
    \Omega_{\rm dS}\hat{\mathbf L}_{\rm o}\times\hat{\mathbf L}_{\rm i}.
\end{equation}
The corresponding precession frequency is
\begin{equation}
    \Omega_{\rm dS}
    \simeq
    \frac{3}{2}
    \frac{m_3}{a_{\rm o}(1-e_{\rm o}^2)}
    \sqrt{m_3/a_{\rm o}^3}.
\end{equation}
The precession of $\hat{\mathbf L}_{\rm i}$ makes the polarization angle $\psi$ time dependent and introduces both the Thomas phase $\phi_T$ and the polarization phase $\phi_p$.
The inclination angle between $\hat{\mathbf L}_{\rm i}$ and the source direction $\hat{\mathbf N}(\Phi,\Theta)$ changes as well, so the polarization amplitudes
\begin{equation}
    A_+=\frac{1+\left(\hat{\mathbf L}_{\rm i}\cdot\hat{\mathbf N}\right)^2}{2},
    \qquad
    A_\times=\hat{\mathbf L}_{\rm i}\cdot\hat{\mathbf N}
\end{equation}
evolve during the observation.

For a space-based detector, the antenna response functions $F_+$ and $F_\times$ project the two GW polarizations onto the detector output~\cite{Space_based_detector_response}.
The detected strain is therefore
\begin{equation}
    h=\sqrt{F_+^2 A_+^2+F_\times^2 A_\times^2} 
    h_0 {\rm e}^{-i(\phi_p+\phi_D)}.
\end{equation}
The annual Doppler phase caused by the detector motion is
\begin{equation}
    \phi_D=2\pi f {\rm A.U.}\times\sin\Theta\cos(2\pi t/{\rm yr}-\Phi),
\end{equation}
and the polarization phase is
\begin{equation}
    \phi_p=\arctan\left[(A_\times F_\times)/(A_+F_+)\right].
\end{equation}
The GW signal considered here is the inspiral signal of the inner BBH.
The GW emitted by the BBH center-of-mass orbit around the SMBH lies at a much lower frequency, typically $\sim10^{-6} {\rm Hz}$ for the systems considered here~\cite{my_paper_smbh}.
We therefore neglect this low-frequency component.
Because laser frequency noise is not modeled in the present calculation, we do not implement time-delay interferometry (TDI).
We nevertheless include the detector transfer function in $F_+$ and $F_\times$ to keep the response close to that required for future space-based observations~\cite{my_paper_polarization_constraints}.
For calculations that include only instrumental and foreground noise, including or omitting TDI has little effect~\cite{my_paper_TDI_combinations}.
This approximation therefore provides a practical balance between model completeness and computational cost.
A complete detector analysis would additionally require realistic TDI response and noise modeling~\cite{Du_Taiji_pipeline_2026}.

The orbital motion of the BBH center of mass around the SMBH requires a moving-source transformation.
Using the Lorentz transformation of the three-dimensional GW tensor, we transform the rest-frame waveform to the frame in which the detector receives radiation from a moving source~\cite{Moving_source_waveform}.
In complex-strain notation, a pure Lorentz transformation gives an additional factor ${\rm e}^{-2i\lambda}$, where $\lambda$ is the boost-induced phase~\cite{Moving_source_waveform}.
This factor depends on the propagation direction and the source velocity, not on the detailed waveform amplitude.
For uniform source motion, $\lambda$ is time independent and can be absorbed into the initial GW phase.
For the hierarchical system considered here, the source velocity varies along the outer orbit.
$\lambda(t)$ must therefore be retained as part of the moving-source waveform transformation.
The boost also changes the apparent propagation direction through aberration.
Consequently, the angles $\Theta$ and $\Phi$ in the detector response become time-dependent quantities, $\Theta(t)$ and $\Phi(t)$.
The amplitude and response factors in the detected strain then acquire additional modulation from the outer-orbit velocity.

Beyond the phase and response modulations described above, the moving-source transformation also relates source and detector times. Let $\tau_S$ be the proper time of the BBH center of mass and $t_D$ the time measured by a stationary detector at infinity. An observer with four-velocity $u^\mu$ measures the frequency $\omega=-k_\mu u^\mu$, where $k^\mu$ is the GW wave vector. Phase invariance then yields~\cite{GR_textbook,GW190521_LISA_orbit_environment}
\begin{equation}
    \frac{{\rm d}\tau_S}{{\rm d}t_D}
    =
    \frac{\omega_D}{\omega_S}
    =
    \frac{k_{\mu,D}u_{D}^{\mu}}{k_{\mu,S}u_{S}^{\mu}}.
    \label{eq:covariant_time_mapping}
\end{equation}
At each emission event, we introduce a local static orthonormal frame outside the ergosphere. In this frame, the source four-velocity and GW wave vector take the standard local forms~\cite{GR_textbook,Moving_source_waveform}
\begin{equation}
    u_S
    =
    \gamma\left(1,\mathbf v\right),
    \qquad
    k_S
    =
    \omega_{\rm stat}\left(1,\hat{\mathbf N}\right),
    \label{eq:local_source_wavevector}
\end{equation}
where $\omega_{\rm stat}$ is the frequency measured by the local static observer, $\mathbf v$ is the BBH center-of-mass velocity, $\hat{\mathbf N}$ is the local GW propagation direction, and $\gamma=(1-v^2)^{-1/2}$. Their Minkowski scalar product gives
\begin{equation}
    \omega_S
    =
    \gamma\left(1-\mathbf v\cdot\hat{\mathbf N}\right)
    \omega_{\rm stat}.
    \label{eq:local_doppler_mapping}
\end{equation}

Thus, $1-\mathbf v\cdot\hat{\mathbf N}$ is the longitudinal Doppler factor, while $\gamma$ accounts for transverse-Doppler time dilation~\cite{Moving_source_waveform,SMBH_triple_transverse_Doppler}. In stationary Kerr spacetime, $E\equiv-k_t$ is the conserved Killing energy of the GW wave vector. Evaluating $g_{tt}$ at emission and using $g_{tt}^{\infty}=-1$ for the stationary detector gives the stationary-observer frequencies~\cite{Kerr_metric,Boyer_Lindquist_coordinates}
\begin{equation}
    \omega_{\rm stat}
    =
    \frac{E}{\sqrt{-g_{tt}}},
    \qquad
    \omega_D
    =
    \frac{E}{\sqrt{-g_{tt}^{\infty}}}
    =E.
    \label{eq:static_gravitational_redshift}
\end{equation}
Combining the covariant frequency ratio with the local Doppler and static gravitational-redshift relations yields the time mapping used in our calculation,
\begin{equation}
    \frac{{\rm d}\tau_S}{{\rm d}t_D}
    =
    \frac{\sqrt{-g_{tt}}}
    {\gamma\left(1-\mathbf v\cdot\hat{\mathbf N}\right)}
    =
    \frac{\sqrt{1-2m_3/r_{\rm o}}}
    {\gamma\left(1-\mathbf v\cdot\hat{\mathbf N}\right)}.
    \label{eq:total_time_mapping}
\end{equation}
This mapping determines how the detector samples the rest-frame inspiral phase and couples the relativistic outer motion to the observed waveform phase.
On the Kerr equatorial plane, the gravitational factor in Eq.~(\ref{eq:total_time_mapping}) takes the Schwarzschild form~\cite{Boyer_Lindquist_coordinates}. Equation~(\ref{eq:total_time_mapping}) is used without expansion in the numerical waveform calculation. Its weak-field, low-velocity expansion separates the physical contributions,
\begin{equation}
    \frac{{\rm d}\tau_S}{{\rm d}t_D}
    \simeq
    1+\mathbf v\cdot\hat{\mathbf N}
    +\left(\mathbf v\cdot\hat{\mathbf N}\right)^2
    -\frac{v^2}{2}
    -\frac{m_3}{r_{\rm o}}
    +\mathcal{O}\!\left(v^3,\frac{m_3}{r_{\rm o}}v,\frac{m_3^2}{r_{\rm o}^2}\right).
    \label{eq:time_mapping_expansion}
\end{equation}
The terms involving $\mathbf v\cdot\hat{\mathbf N}$ arise from the longitudinal Doppler effect, $-v^2/2$ is the transverse-Doppler time-dilation term, and $-m_3/r_{\rm o}$ is the gravitational-redshift term. The detector-frame waveform therefore includes longitudinal and transverse Doppler effects, SMBH gravitational redshift, and cosmological redshift through the redshifted masses $m_{i,z}=(1+z)m_i$ in $h_0$.

Our waveform model does not include additional GW propagation effects in the SMBH spacetime, including Shapiro delay and GW lensing. Away from source--SMBH--observer alignment, the characteristic Shapiro delay is $\Delta t_{\rm Sh}\sim 2m_3\ll a_{\rm o}$ and is subdominant to the source-motion effects considered here~\cite{SMBH_triple_direct_measurement,GW190521_LISA_orbit_environment}. The transverse-Doppler and gravitational-redshift contributions retained in our model also exceed the Shapiro contribution across a broad region of the SMBH--BBH parameter space~\cite{SMBH_triple_transverse_Doppler}. Significant GW lensing requires the BBH--SMBH projected separation to be comparable to or smaller than the Einstein radius, and hence is concentrated near alignment~\cite{SMBH_triple_lensing,GW190521_LISA_orbit_environment}. These propagation effects are therefore treated as outside the scope of the present baseline waveform model rather than as identically zero. The approximation is most reliable for generic non-aligned, weak-deflection configurations, while sources with small projected separations or close conjunctions require an explicit treatment. Other propagation corrections, including gravitomagnetic time delays, gravitational Faraday rotation, and the gravitational spin Hall effect, enter at higher order than the Shapiro delay in the non-aligned weak-deflection regime considered here~\cite{SMBH_propagation_frame_dragging,GW_gravitational_Faraday_rotation,GW_spin_Hall_lensing}. We therefore do not include these effects in the present waveform.

Combining the factors above gives the schematic detector-frame GW signal
\begin{equation}
\begin{aligned}
    h(t_D)&=\sqrt{F_+^2 A_+^2+F_\times^2 A_\times^2}h_0(\tau_S)\\
    &\quad\times {\rm e}^{-i(\phi_p+\phi_D+\phi_{\rm DF}+2\phi_T+2\lambda )}.
\end{aligned}
\end{equation}
The rest-frame inspiral waveform is therefore supplemented by the DF phase correction, dS orientation modulation, detector Doppler and polarization phases, moving-source boost factor, and time transformation.
The complete waveform-building flow is summarized in Fig.~\ref{fig:AGN}.

\begin{figure}[!ht]
    \centering
    \includegraphics[width=0.97\columnwidth]{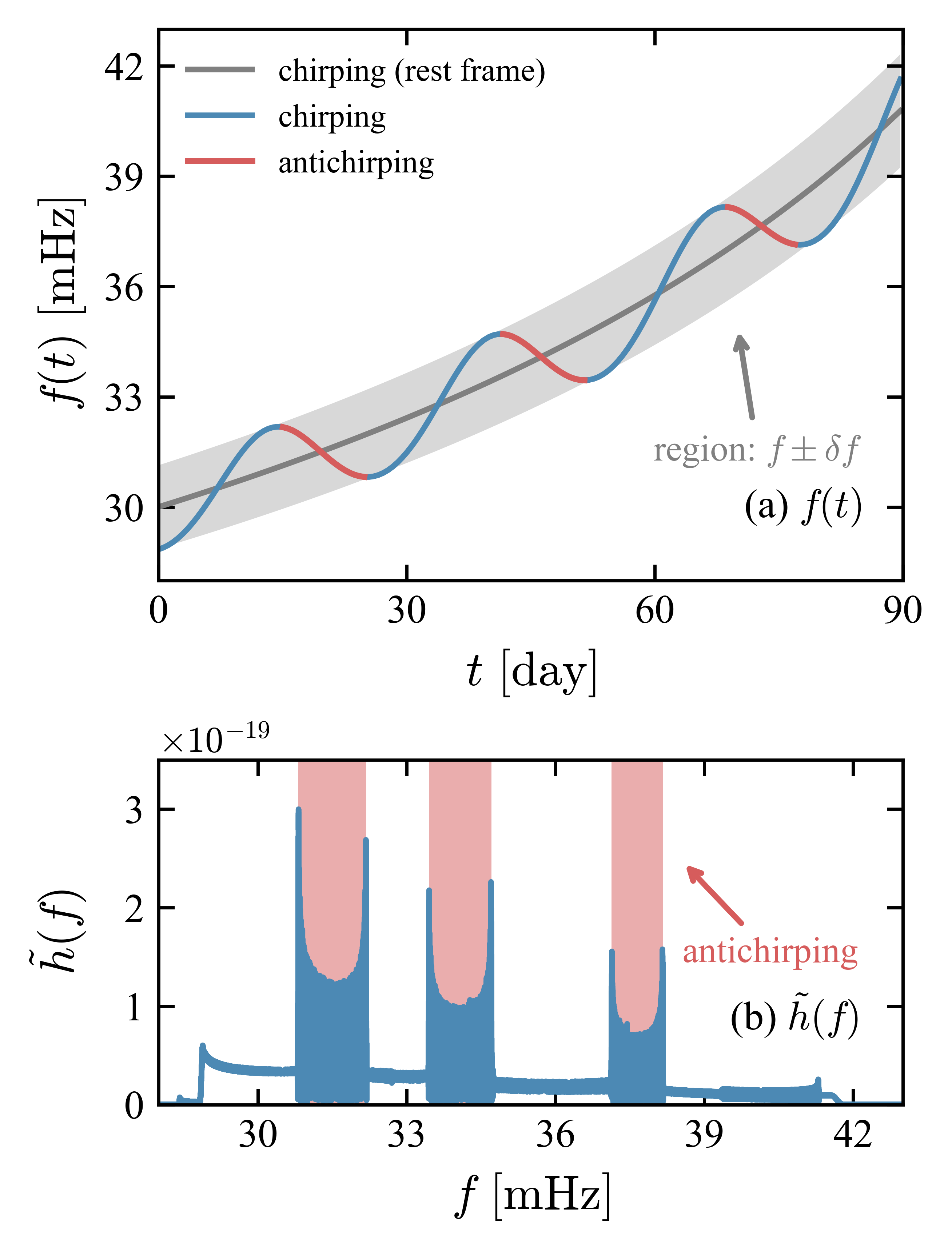}
    \caption{
    Antichirping caused by source motion.
    Panel (a) shows the GW frequency as a function of time.
    The gray region marks the Doppler-shift range $\pm\delta f$, with $\delta f\simeq f\sqrt{m_3/a_{\rm o}}$.
    Panel (b) shows the corresponding frequency-domain GW signal.
    The red region denotes the antichirping part associated with the red curve in panel (a).
    The parameters are $f_0=0.03\,{\rm Hz}$, $m_1=m_2=50\,M_\odot$, $D_L=50\,{\rm Mpc}$, $m_3=4\times10^6\,M_\odot$, $a_{\rm o}=350\,R_s$, and $e_{\rm o}= \chi_3=0$.
    }
    \label{fig:antichirping}
\end{figure}

The periodic motion of the BBH around the SMBH modulates the observed GW frequency through redshift and blueshift~\cite{GW190521_LISA_environment,my_paper_kick}.
This behavior is visible in Fig.~\ref{fig:antichirping}(a).
Relative to the monotonic chirping in the rest frame, the moving-source frequency shows alternating chirping and antichirping intervals.
The originally monotonic rest-frame chirping becomes nonmonotonic under Doppler modulation.
A global stationary phase approximation (SPA) therefore cannot be applied directly.
In Fig.~\ref{fig:antichirping}(b), the chirping and antichirping contributions overlap in frequency space.
This overlap affects SPA-based frequency-domain waveform constructions.
One possible treatment is to segment the time-domain signal into monotonic frequency branches.
The SPA can then be applied to each branch separately when constructing a frequency-domain waveform~\cite{GW190521_LISA_orbit_environment}.
We construct the waveform in the time domain and then Fourier transform it for the SNR and FIM calculations.
The Doppler modulation therefore enters naturally, without the ambiguity that appears when the SPA treats overlapping chirping and antichirping branches.

The complete waveform model combines the hierarchical dynamics in Sec.~\ref{sec:triple_system} with the signal model developed in this section.
The outer orbit and AGN disk determine $\mathbf r_{\rm o}$, $\mathbf v$, and the accumulated DF-induced orbital modulation.
The inner BBH provides the rest-frame inspiral waveform with PN phase, DF phase correction, and dS-induced orientation modulation.
The detector response, moving-source transformation, and time transformation map these ingredients into the observed GW signal.
This construction gives the inspiral GW signal for a BBH embedded in an AGN disk.

\section{Methodology}\label{sec:methodology}

The waveform model above contains the physical effects used in the parameter-estimation analysis of the AGN-disk system.
For a LISA observation with $T_{\rm obs}=1\,{\rm yr}$, we quantify the signal strength and parameter uncertainties using the standard noise-weighted inner product.
It is defined as~\cite{LISA_angular_resolution}
\begin{equation}
    (a|b)
    =
    4 {\rm Re}
    \int
    \frac{\tilde a^*(f)\tilde b(f)}{S_n(f)}
    {\rm d}f,
    \label{eq:inner_product}
\end{equation}
where $\tilde a(f)$ and $\tilde b(f)$ are Fourier transforms of $a(t)$ and $b(t)$, and $S_n(f)$ is the one-sided detector noise power spectral density (PSD).
We adopt the LISA PSD from Ref.~\cite{LISA_noise_PSD} and include the foreground noise formed by Galactic binaries~\cite{Galactic_white_dwarf_foreground,Compact_binary_stochastic_LISA,my_paper_galactic_confusion_foreground}.
We restrict our analysis to the BBH signal and the foreground noise, neglecting overlaps with other GW signals~\cite{Ma_GCB_EMRI_2026}.

The signal-to-noise ratio is
\begin{equation}
    {\rm SNR}=\sqrt{(h|h)}.
    \label{eq:snr}
\end{equation}
We exclude GW signals with ${\rm SNR}<8$ from the parameter-estimation sample. We adopt ${\rm SNR}\geq8$ as the reference threshold to ensure detectable signals and resolvable gas density (see Sec.~\ref{subsec:results_parameter_impact}). Although a higher threshold improves the typical precision, it also reduces the sample size. We therefore use this fixed threshold to define a consistent population for the subsequent FIM analysis.

We use the FIM to estimate parameter uncertainties~\cite{Fisher_matrix}.
For parameters $\boldsymbol{\theta}$, the FIM is
\begin{equation}
    \Gamma_{ij}
    =
    \left(
    \frac{\partial h}{\partial\theta_i}
    \middle|
    \frac{\partial h}{\partial\theta_j}
    \right).
    \label{eq:fisher_matrix}
\end{equation}
The 16 parameters varied in the FIM calculation are
\begin{equation}
    \begin{aligned}
    \boldsymbol{\theta}
    =
    \{&
    \ln f_0,\ln m_1,\ln m_2,\ln D_L,\Theta,\Phi,\Psi,
    \\
    &
    \ln m_3,\ln a_{\rm o},e_{\rm o},\chi_3,
    \epsilon,\zeta,\lambda_L,\ln\rho_0,\alpha
    \}.
    \end{aligned}
    \label{eq:fisher_parameters}
\end{equation}
The angle $\lambda_L$ denotes the angle between the outer and inner orbital angular momenta, with $\cos\lambda_L=\hat{\mathbf L}_{\rm o}\cdot\hat{\mathbf L}_{\rm i}$.
The angles $\epsilon$ and $\zeta$ connect the SMBH coordinates to the Solar System barycentric coordinates, following the coordinate transformation given in the Appendix of our previous work~\cite{my_paper_smbh}.
For positive parameters with broad dynamic ranges, logarithmic variables reduce the numerical ill conditioning that can arise when the corresponding linear variables are used directly in the FIM.
The inverse matrix $\Sigma=\Gamma^{-1}$ is the covariance matrix.
The one-sigma uncertainty of $\theta_i$ is
\begin{equation}
    \sigma_{\theta_i}=\sqrt{\Sigma_{ii}}.
    \label{eq:parameter_uncertainty}
\end{equation}
We check numerical derivatives by repeating the calculation with multiple finite-difference step sizes.
We select steps for which both the derivatives and the resulting uncertainties reach a stable plateau.
The typical finite-difference step is $10^{-6}$ for linear parameters.
For logarithmic parameters, the typical relative perturbation of the original positive quantity is $10^{-8}$.
We follow the matrix-inversion strategy in Ref.~\cite{FIM_matrix_inversion} and monitor the residual
\begin{equation}
    \epsilon_{\rm inv}
    =
    \max_{ij}
    \left|
    \left(\Gamma\Sigma-I\right)_{ij}
    \right|.
    \label{eq:inversion_residual}
\end{equation}
We retain only calculations satisfying $\epsilon_{\rm inv}<10^{-4}$.
This procedure helps ensure the numerical reliability of the FIM calculation.

We also estimate systematic parameter biases caused by differences between waveform models.
The reference signal $h_{\rm DF}$ is the full waveform and is treated as the true signal in this bias calculation.
The comparison waveform $h$ removes both DF effects.
In this waveform, the outer-orbit acceleration does not include $\mathbf a_{\rm DF}$, and the intrinsic phase does not include $\phi_{\rm DF}$.
All physical parameters are otherwise kept identical in the two waveforms.
The residual is $\delta h=h_{\rm DF}-h$.
The induced systematic error is~\cite{waveform_systematic_error}
\begin{equation}
    \Delta\theta_i
    =
    \sum_j
    \left(\Gamma^{-1}\right)_{ij}
    \left(
    \frac{\partial h}{\partial\theta_j}
    \middle|
    \delta h
    \right).
    \label{eq:systematic_bias}
\end{equation}
This expression quantifies how gas-induced waveform differences can bias BBH, SMBH, and outer-orbit parameters when the comparison waveform omits the environmental contribution.
Only the FIM used for the systematic-error calculation is constructed from the comparison waveform and excludes $\rho_0$ and $\alpha$.
All other FIM calculations use the full waveform and include the AGN disk parameters.

\begin{table}[!ht]
\centering
\footnotesize
\renewcommand{\arraystretch}{1.4}
\setlength{\tabcolsep}{3pt}
\caption{
Parameter distributions used in this work, following Ref.~\cite{LISA_stellar_origin_background}.
$U[a,b]$ denotes a uniform distribution from $a$ to $b$.
}
\label{tab:method_parameters}
\begin{tabular}{p{0.46\columnwidth}p{0.47\columnwidth}}
\hline
Parameter & Distribution \\
\hline
$m$ [$M_\odot$] & $U[20,230]$ \\
$q_{\rm i}$ & $U[0.3,1]$ \\
$\chi_{1z},\chi_{2z}$ & $U[-0.9,0.9]$ \\
$\Phi,\Psi,\zeta$ [rad] & $U[0,2\pi]$ \\
$\Theta,\epsilon,\lambda_L$ [rad] & $\arccos\!\left(U[-1,1]\right)$ \\
$\log_{10}(z)$ & $U[-2.17,-0.04]$ \\
$\tau$ [yr] & $U[10,1000]/(1+z)$ \\
$\log_{10}(m_3)$ [$M_\odot$] & $U[5,9]$ \\
$\log_{10}(a_{\rm o})$ [$R_s$] & $U[1.47,3.47]$ \\
$\chi_3$ & $U[0,0.95]$ \\
$e_{\rm o}$ & $U[0,0.95]$ \\
$\alpha$ & $U[0.8,3.2]$ \\
$\log_{10}(\rho_0)$ [{\rm g\,cm}$^{-3}$] & $U[-12.5,-8.5]$ \\
\hline
\end{tabular}
\end{table}

We obtain the component masses $m_1$ and $m_2$ from the sampled total mass $m=m_1+m_2$ and inner mass ratio $q_{\rm i}=m_2/m_1$.
The initial GW frequency $f_0$ follows from $\tau$ as~\cite{my_paper_PN_models}
\begin{equation}
    f_0
    =
    \frac{1}{8\pi}
    \left(
    \frac{1}{5}M_c^{5/3}\tau
    \right)^{-3/8}.
    \label{eq:initial_frequency}
\end{equation}
The luminosity distance $D_L$ is computed from $z$ through
\begin{equation}
    D_L
    =
    \frac{1+z}{H_0}
    \int_0^z
    \frac{{\rm d}z'}
    {\sqrt{\Omega_m(1+z')^3+\Omega_\Lambda}}.
    \label{eq:luminosity_distance}
\end{equation}
We adopt the Lambda cold dark matter cosmological model with parameters from the \textit{Planck 2018 results}~\cite{Planck_2018}: $H_0=67.37\,{\rm km\,s^{-1}\,Mpc^{-1}}$, $\Omega_m=0.315$, and $\Omega_\Lambda=0.685$.
For each sampled system, we integrate the outer orbit with the relativistic and DF acceleration terms described in Sec.~\ref{sec:triple_system}.
We then generate the detector-frame waveform according to Sec.~\ref{sec:gw_signal} and evaluate the FIM from numerical waveform derivatives.
The time-domain signal is sampled at five times the GW frequency at the end of the observation.
This rate is above the Nyquist requirement and preserves the waveform information used in the Fourier transform.
In total, we generate 1600 hierarchical triple systems using the parameter distributions in Table~\ref{tab:method_parameters}.
After removing systems with ${\rm SNR}<8$ or failed inversion-residual tests, about 200 valid systems remain for the parameter-estimation analysis.

\section{Results}\label{sec:results}

\subsection{Impact of the AGN disk on parameter estimation}\label{subsec:results_parameter_impact}
\begin{figure}[!ht]
    \centering
    \includegraphics[width=0.98\columnwidth]{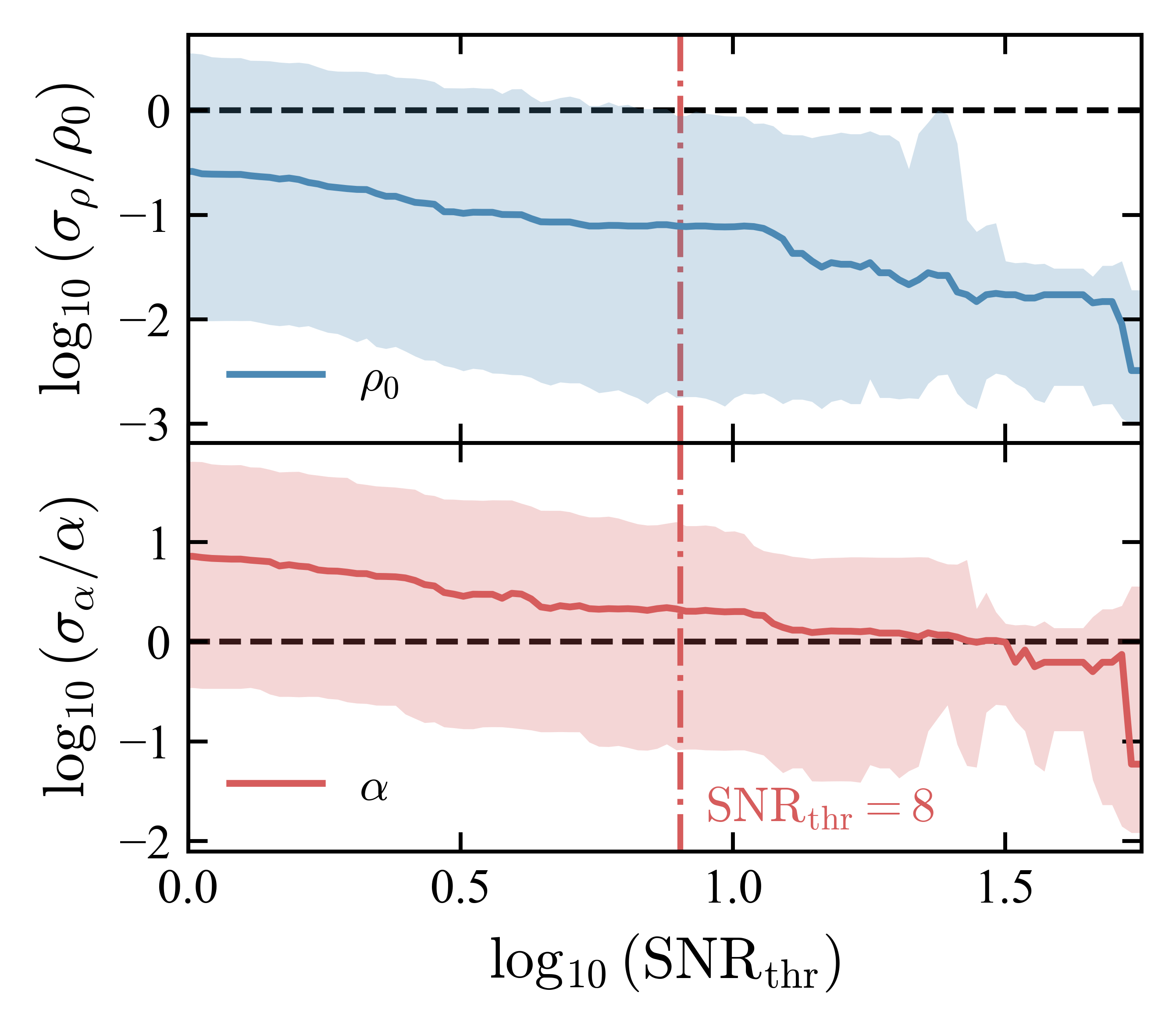}
    \caption{
Relative uncertainties in the AGN disk parameters versus the SNR threshold. The upper and lower panels show $\rho_0$ and $\alpha$, respectively. For each ${\rm SNR}_{\rm thr}$, the population includes all valid systems with ${\rm SNR}\geq{\rm SNR}_{\rm thr}$. Solid curves show median relative uncertainties, and shaded regions span the 16th--84th percentiles. The vertical red line marks ${\rm SNR}_{\rm thr}=8$. Horizontal black dashed lines mark a relative uncertainty of $1$, which provides an approximate resolvability threshold.
    }
    \label{fig:snr_threshold}
\end{figure}

For the following population analysis, we adopt ${\rm SNR}\geq8$ as the reference threshold. As quantified in Fig.~\ref{fig:snr_threshold}, this threshold balances the removal of weak signals against the retention of a sufficiently large sample for robust population statistics.

\begin{figure}[!ht]
    \centering
    \includegraphics[width=0.97\columnwidth]{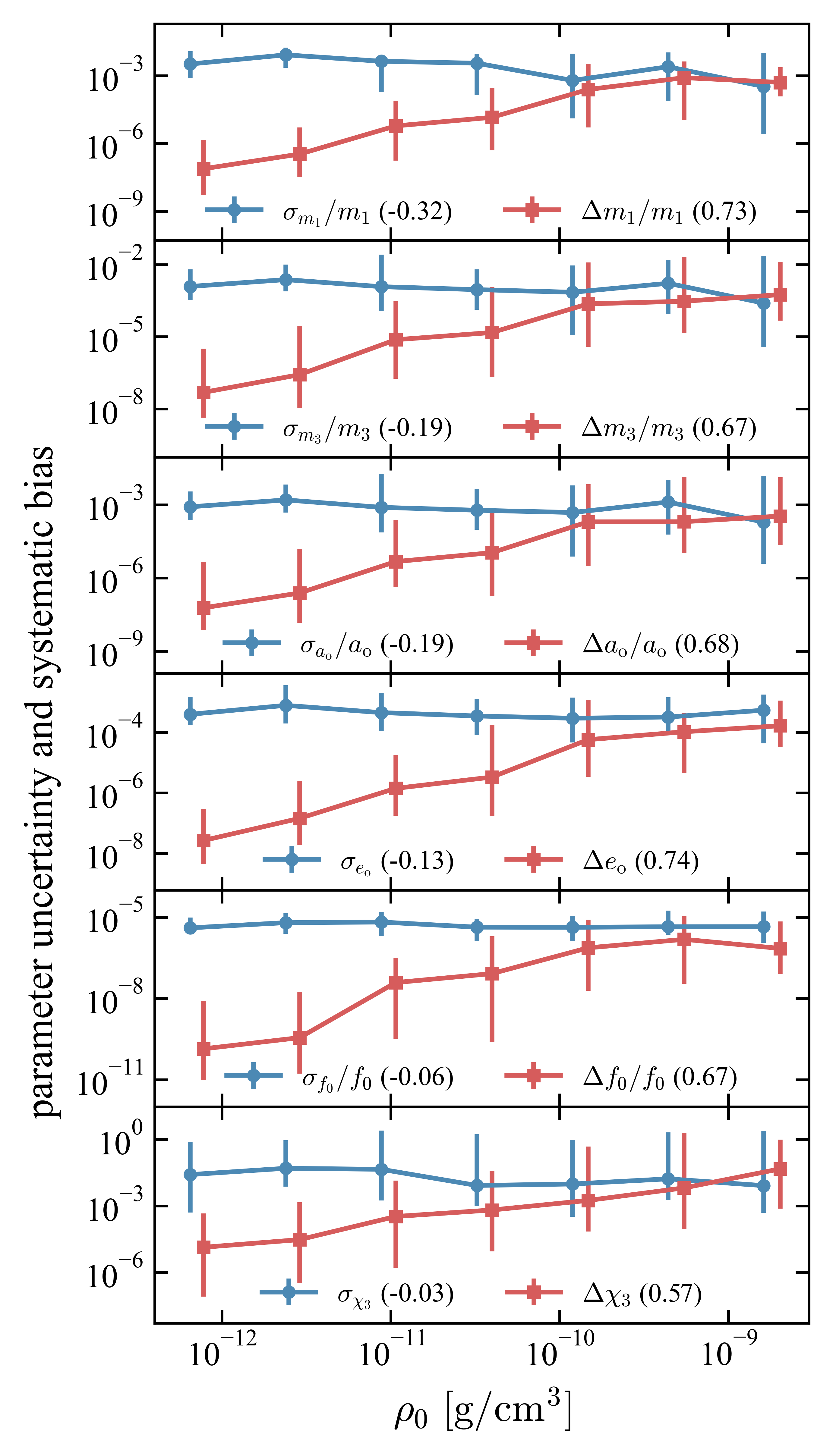}
    \caption{
    Gas-density dependence of parameter uncertainties and systematic biases.
    Blue markers show the Fisher parameter uncertainties $\sigma_\theta$, and red markers show the systematic biases $\Delta\theta$.
The panels show representative parameters correlated with $\rho_0$, including $m_1$, $m_3$, $a_{\rm o}$, $e_{\rm o}$, $f_0$, and $\chi_3$.
    The numbers in parentheses are Pearson correlation coefficients computed in logarithmic space.
    The markers and error bars are obtained by binning the horizontal variable and computing the median and the 16th and 84th percentiles in each bin.
    }
    \label{fig:rho0_pars}
\end{figure}

Figure~\ref{fig:snr_threshold} shows that relative uncertainties decrease as the SNR threshold increases. The median uncertainty in $\rho_0$ remains below 1, and its $1\sigma$ interval lies below 1 for ${\rm SNR}_{\rm thr}\geq8$. This supports our choice of ${\rm SNR}\geq8$ as the reference threshold. Systems satisfying this criterion form the sample used in all subsequent figures and discussion. The constraint on $\alpha$ is weaker, as only part of the population reaches a relative uncertainty below 1. These SNR-selected systems are used below to examine how gas density modifies the waveform and affects source-parameter inference.

The disk-parameter uncertainties below are obtained from the full 16-dimensional FIM and are therefore marginalized over all source and disk parameters. The marginalized covariance matrix shows correlations between $\rho_0$ and the binary masses, SMBH mass, and outer semimajor axis, reflecting their effects on the accumulated GW phase and source-motion modulation. These correlations are partial rather than exact parameter degeneracies, as their waveform dependences differ in frequency and outer-orbit structure. To assess their typical impact, we use single-parameter conditional FIM calculations in which $m_1$, $m_3$, or $a_{\rm o}$ is fixed in turn. For most representative sources, fixing one of these parameters reduces the uncertainty in $\rho_0$ by no more than approximately $20\%$. This is not an order-of-magnitude improvement and indicates that the density constraint is not determined solely by these correlations. All retained FIM calculations also satisfy the inversion-residual criterion in Eq.~(\ref{eq:inversion_residual}), indicating that the covariance-matrix inversion is numerically reliable and that no numerically pathological inversion affects the reported uncertainties.

The resulting FIM therefore contains partial parameter degeneracies, but the parameters remain distinguishable through their different waveform dependences. The gas density consequently remains measurable. Complementary observations can further reduce the remaining correlations. Independent ground-based and space-based GW observations can constrain complementary source parameters~\cite{my_paper_smbh,my_paper_kick}, while electromagnetic observations of the AGN environment can provide external information on the SMBH--disk configuration~\cite{GW190521_AGN_counterpart}. Such external information may therefore improve the density measurement.

Gas DF changes the outer orbit and velocity of the BBH center of mass through $\mathbf a_{\rm DF}$.
The same gas environment also changes the intrinsic inspiral phase through $\phi_{\rm DF}$.
Both channels accumulate over the one-year observation, allowing even a perturbative DF acceleration to become visible in the phase and moving-source modulation.

Figure~\ref{fig:rho0_pars} shows the accumulated gas imprint on representative parameter uncertainties and systematic errors.
The parameters in Fig.~\ref{fig:rho0_pars} are selected from the full 16-dimensional FIM according to their correlations with the gas density.
For the same displayed parameters, the correlations with $\alpha$ are negligible for both the statistical uncertainties and systematic errors.
It is therefore not shown here.
As $\rho_0$ increases, statistical uncertainties show weak, parameter-dependent trends, whereas systematic errors increase.
A denser gas disk produces a larger perturbation to the outer motion and a larger accumulated phase correction.
The waveform then becomes more sensitive to the affected parameters, but the bias also grows when the DF contribution is omitted from the comparison model.

Among the displayed parameters, the uncertainty in $m_1$ shows the strongest statistical trend, with a Pearson coefficient of $-0.32$. The remaining uncertainties are only weakly correlated with $\rho_0$, with $r=-0.19$ for both $m_3$ and $a_{\rm o}$, $r=-0.13$ for $e_{\rm o}$, $r=-0.06$ for $f_0$, and $r=-0.03$ for $\chi_3$. Thus, the statistical precision of these source parameters varies weakly with $\rho_0$ and remains configuration dependent.

The red curves show the systematic parameter biases induced by using the comparison waveform in place of $h_{\rm DF}$.
Across the sampled systems, systematic errors grow with $\rho_0$ despite the weak statistical trends.
The gas effect that makes the waveform more informative also makes an incomplete waveform model less accurate.
At sufficiently large density, the red curves can exceed the blue curves for several parameters.
This behavior indicates that the DF correction cannot be treated as a negligible perturbation in parameter inference.

A zero-density uncertainty curve is not shown for the AGN parameters.
If DF is removed from the waveform model, the FIM elements associated with $\partial h/\partial \rho_0$ and $\partial h/\partial \alpha$ are zero.
The inverse FIM is then formally divergent, making the corresponding uncertainty estimates unusable.
If the rows and columns associated with $\rho_0$ and $\alpha$ are removed instead, the calculation is no longer a zero-density comparison in the same parameter space.
Such a calculation is equivalent to assuming that the AGN parameters are fixed a priori.
This assumption artificially reduces the uncertainties of the remaining parameters, as in analogous parameter-fixing treatments in foreground analyses~\cite{Taiji_Galactic_confusion_noise,my_paper_galactic_confusion_foreground}.
For this reason, the impact of omitting DF is characterized here through systematic errors rather than through a reduced-parameter zero-density FIM.

Figure~\ref{fig:rho0_pars} therefore shows two complementary consequences of the AGN disk.
Increasing $\rho_0$ changes the statistical precision of the displayed parameters only weakly and in a parameter-dependent way, but it increases the systematic error caused by an incomplete waveform model.
Once the systematic error becomes comparable to or larger than the Fisher uncertainty, the DF contribution must be included consistently in the waveform.

\subsection{Constraints on AGN disk parameters}\label{subsec:results_agn_constraints}
\begin{figure}[!ht]
    \centering
    \includegraphics[width=0.97\columnwidth]{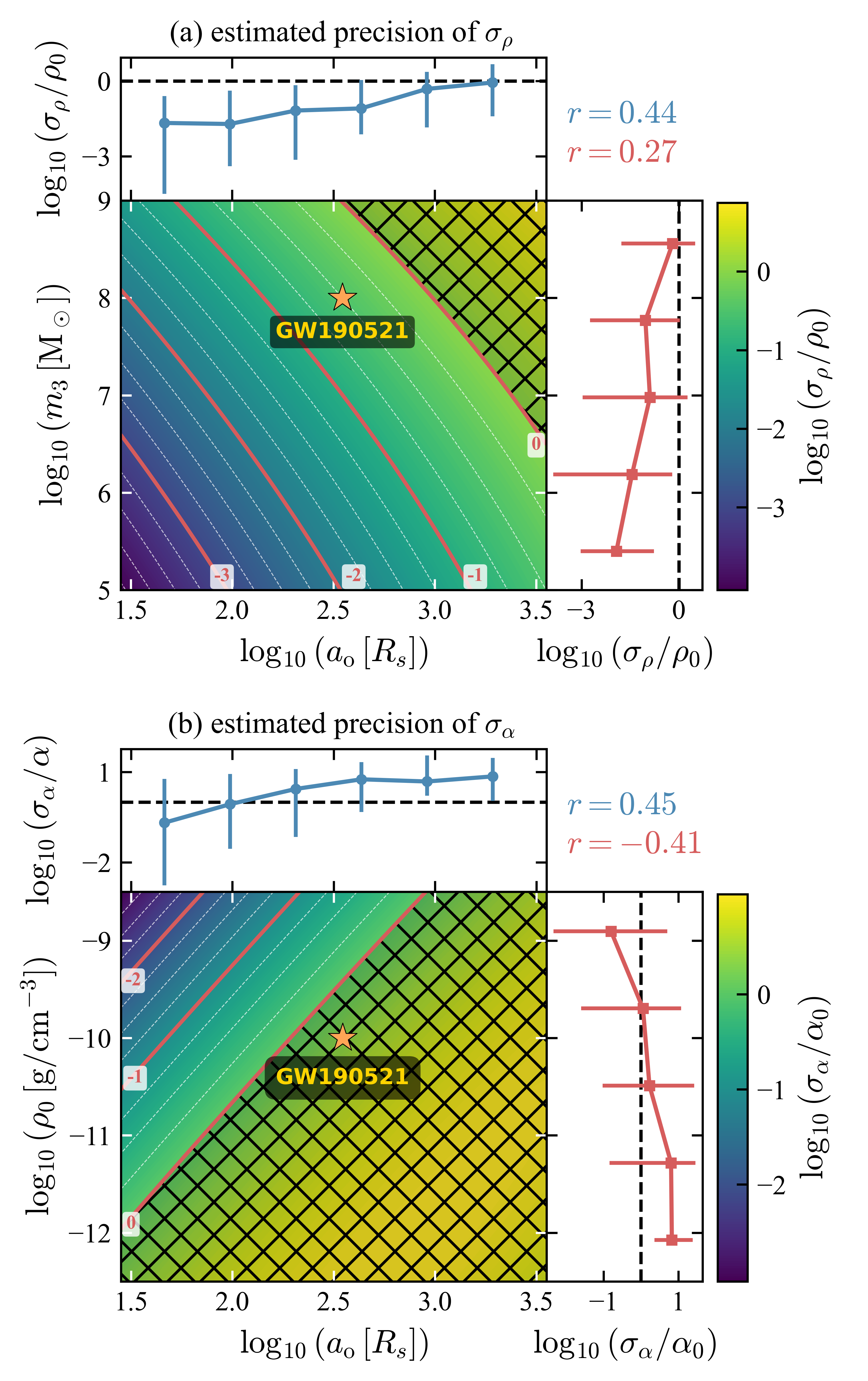}
\caption{Estimated relative precision of the AGN disk density normalization $\rho_0$ and density-profile index $\alpha$.
Panel (a) shows $\sigma_\rho/\rho_0$ as a function of $a_{\rm o}$ and $m_3$, and panel (b) shows $\sigma_\alpha/\alpha$ as a function of $a_{\rm o}$ and $\rho_0$.
The side panels show computed results from the sampled systems, with the same markers and error bars as in Fig.~\ref{fig:rho0_pars}. The black dashed lines mark unit relative uncertainty.
    The central color maps are polynomial fits to the computed Fisher results using $z=c_0+\sum_{i=1}^{2}(a_i x^i+b_i y^i)$.
The variables $x$, $y$, and $z$ are taken in the same linear or logarithmic space as displayed in the corresponding panel.
Cross-hatched regions indicate relative uncertainties larger than unity, for which the corresponding parameter is unresolved.
    The fits are used only as smoothed visual representations of the sampled results and are not used for extrapolation outside the sampled parameter region.
    The $r$ values in the upper-right corners denote Pearson correlation coefficients.
    The star denotes the GW190521-like benchmark with $m_3=10^8\,\rm M_\odot$, $a_{\rm o}=350\,R_s$, and $\rho_0=10^{-10}\,{\rm g\,cm^{-3}}$~\cite{GW190521_AGN_counterpart}.
    }
    \label{fig:estimated_precision_DF}
\end{figure}

We next identify the system parameters that most strongly affect the relative uncertainties of the AGN disk parameters.
We summarize the Fisher constraints on $\rho_0$ and $\alpha$ in Fig.~\ref{fig:estimated_precision_DF}. 
For each AGN parameter, the two displayed axes correspond to the parameters with the largest correlations with the relevant relative uncertainty.
Figure~\ref{fig:estimated_precision_DF} therefore shows how the AGN disk constraints correlate with other system parameters and gives the order-of-magnitude precision that can be achieved.

In Fig.~\ref{fig:estimated_precision_DF}(a), the relative density uncertainty increases with $a_{\rm o}$, with a Pearson coefficient of $0.44$.
The positive trend indicates that the density constraint weakens as the outer semimajor axis increases.
At larger $a_{\rm o}$, the BBH is farther from the SMBH and the gas density is lower.
The DF imprint is therefore weaker.
The Doppler modulation from the outer motion is also reduced.
Changes in $\rho_0$ therefore become less distinguishable in the waveform.
The correlation with $m_3$ is weaker than that with $a_{\rm o}$, with $r=0.27$, but it remains one of the strongest correlations for $\sigma_\rho$ among the sampled parameters.
As $m_3$ increases, the outer orbital period becomes longer and the relative change in orbital velocity becomes smaller.
These effects modify the accumulation of the source-motion modulation in the observed signal.
For most sampled systems, GW observations constrain the density accurately.

In Fig.~\ref{fig:estimated_precision_DF}(b), the relative uncertainty in $\alpha$ is most strongly correlated with $a_{\rm o}$, with $r=0.45$.
The uncertainty increases with $a_{\rm o}$ for the same physical reason as in the density measurement.
At larger orbital radii, the local gas density is lower and the DF effect is weaker.
The Doppler modulation is also less effective in encoding the radial structure of the disk.
The second strongest trend is with $\rho_0$, with $r=-0.41$.
The negative sign indicates that a denser disk gives a tighter constraint on $\alpha$.
This follows from the role of $\alpha$ in controlling the radial variation of the DF strength.
If the overall gas density is small, the radial variation is difficult to separate from other waveform modulations.
As $\rho_0$ increases, the radial dependence governed by $\alpha$ leaves a stronger imprint on both the accumulated phase and the outer-orbit modulation.
Overall, constraints on $\alpha$ are weaker. Only systems with high $\rho_0$ and small $a_{\rm o}$ can potentially constrain $\alpha$ accurately. Precise GW-only measurements of $\alpha$ therefore remain challenging.

GW190521 has been discussed as a possible BBH merger in an AGN environment, although this interpretation remains debated~\cite{GW190521_AGN_counterpart,GW190521_AGN_disk_origin,GW190521_AGN_association_insufficient,GW190521_AGN_flare_confidence}.
Here the GW190521-like point is used only as a representative configuration.
If a similar system entered the frequency band of a space-based detector several years before merger, the forecast density uncertainty spans $\sigma_\rho\sim10^{-12}\text{--}10^{-10}\,{\rm g\,cm^{-3}}$ under the adopted LISA configuration and a one-year observation. The parameter $\alpha$ is generally not constrained with useful precision.
The present GW-only forecasts therefore constrain $\rho_0$ more effectively than $\alpha$.
If theoretical studies or astrophysical observations constrain the disk density profile and fix $\alpha$, the median uncertainty in $\rho_0$ is reduced by $12.1\%$. Forecasts that vary both $\rho_0$ and $\alpha$ are therefore conservative.

Figures~\ref{fig:rho0_pars} and~\ref{fig:estimated_precision_DF} show that the gas density affects both source-parameter inference and the measurability of the disk parameters.
In summary, these results show two connected aspects of the same environmental effect.
The DF contribution can improve the statistical measurement of several BBH, SMBH, and outer-orbit parameters, but it also produces systematic errors if omitted from the waveform.
When $\rho_0$ and $\alpha$ are treated as target parameters, the forecasts indicate that space-based observations of favorable hierarchical BBHs can constrain the gas density in AGN disks directly.

\section{Conclusions}\label{sec:conclusions}

In this paper, we investigated whether GWs from inspiraling stellar-mass BBHs in hierarchical triple systems can constrain AGN disk properties.
The system consists of a BBH orbiting a Kerr SMBH inside an AGN disk.
The outer orbit is evolved with relativistic orbital terms and gaseous DF.
The waveform model includes the rest-frame BBH inspiral, dS precession, the DF phase correction, the moving-source transformation, and the detector response.
We then used the FIM to calculate parameter uncertainties and systematic errors for one-year LISA observations of the sampled hierarchical triples.

Our results indicate that AGN gas can leave measurable imprints on both the source motion and the intrinsic BBH phase.
As the gas density increases, the statistical uncertainties of several BBH, SMBH, and outer-orbit parameters show weak, parameter-dependent trends.
At the same time, the systematic errors caused by omitting DF increase.
The gas contribution should therefore be included consistently once the environmental imprint becomes comparable to the statistical uncertainty.
When the AGN disk parameters are treated as target parameters, forecasts for favorable GW190521-like systems reach $\sigma_\rho\sim10^{-12}\text{--}10^{-10}\,{\rm g\,cm^{-3}}$. The parameter $\alpha$ remains difficult to measure accurately with GW observations alone under the assumptions adopted here.

Long-duration space-based GW observations therefore provide a possible way to extract environmental information from hierarchical BBH inspirals.
Such systems can act as dynamical probes of AGN disks because the observed waveform carries both intrinsic inspiral information and source-motion modulation.
The scope of this calculation is set by several controlled assumptions.
These assumptions include the Fisher approximation, a simplified power-law AGN disk, Keplerian gas flow, a local DF prescription, and a one-year LISA observation.
Within this scope, moving BBH sources carry additional environmental information.
The resulting forecasts should therefore be interpreted within the locally Gaussian likelihood assumption of the Fisher framework and the physical assumptions listed above.

Future work will extend the present analysis in several directions. A full Bayesian Markov Chain Monte Carlo analysis will test the Fisher forecasts and assess possible non-Gaussian posterior structure.
Future work will also combine the full LK secular dynamics, relativistic precession, and GW radiation reaction with strong-field Kerr propagation to assess the sources that fall outside the baseline approximation.
Taiji, TianQin, and space-based detector networks can be included to assess the improvement from multiple detector configurations and realistic data-analysis pipelines.
Possible electromagnetic observations of the AGN host can also be combined with space-based inspiral observations and ground-based merger observations.
Advanced denoising methods can improve the recovery of ground-based merger signals~\cite{Ma_ET_denoising_2025}, while eccentric higher harmonics can aid early detection and localization~\cite{Yang_eccentric_harmonics_2026}.
Such a multi-observation strategy may further improve parameter constraints.
Future global-fit analyses can also incorporate methods that combine deep-learning signal separation with Bayesian inference to address overlapping GW signals~\cite{Ma_DL_Bayesian_2025}.
For special viewing geometries, gravitational-lensing effects caused by the SMBH may affect the observed signal and should be included in a more complete analysis.
These developments will provide a more complete assessment of how space-based GW observations can probe AGN disk structure through hierarchical BBH inspirals.

\begin{acknowledgments}
This work was supported by the National Key Research and Development Program of China (Grant No. 2023YFC2206702), the National Natural Science Foundation of China (Grants No. 125B2102, No. 12575072, and No. 12547101), the Fundamental Research Funds for the Central Universities Project (Grant No. 2024IAIS-ZD009), and the Natural Science Foundation of Chongqing (Grant No. CSTB2023NSCQ-MSX0103).
\end{acknowledgments}

\bibliographystyle{unsrt}
\bibliography{references}
\end{document}